\documentclass[preprint,12pt,authoryear]{elsarticle}
\usepackage{amssymb}
\usepackage{graphicx}
\usepackage{lineno}

\journal{Physica A}

\begin{document}

\begin{frontmatter}

\title{Activity delay patterns in project networks}

\author{Alexei Vazquez\corref{cor1}, Chrysostomos Marasinou, Georgios Kalogridis and
Christos Ellinas\corref{cor2}}
\affiliation{Nodes \& Links Ltd, Salisbury House, Station Road, Cambridge, CB1 2LA, UK}
\cortext[cor1]{alexei@nodeslinks.com}
\cortext[cor2]{christos@nodeslinks.com}

\begin{abstract}
Delays in activities completion drive human projects to schedule and cost overruns. It is believed activity delays are the consequence of multiple idiosyncrasies without specific patterns or rules. Here we show that is not the case. Using data for 180 construction project schedules, we demonstrate that activity delays satisfy a universal model that we call the law of activity delays. After we correct for delay risk factors, what remains follows a log-normal distribution.
\end{abstract}

\begin{keyword}
project networks, activity delays, delay risk factors, lognormal distribution; natural language processing; complexity
\end{keyword}

\end{frontmatter}

%\linenumbers

\section{Introduction}
\label{sec:introduction}

We are planners. We plan our day and our lives. We plan at home and at work. We break down plans into discrete activities and aggregate activities into projects. Projects account for 20 and 50\% of economic activity depending on the country \cite{scranton14, jensen16}. However, projects rarely progress as planned.

About 75\% of construction projects are delayed, with a median delay between 20 and 40\% of the project duration depending on the sector \cite{rivera16, budzier19, park21, natarajan22}. Based on data for eight software engineering projects, 20\% of activities are late, with delays between one to hundred of days \cite{choetkiertikul17}. NASA space projects are delayed by 20\% of project duration, with maximum schedule growth near 85\% \cite{emmons07}. Over 75\% of crowdfunded projects deliver the intended product later than planned, with delays of up to ten months \cite{mollick14}. PhD completion takes on average ten months
longer than expected, sometimes taking five years extra \cite{schoot13}. The duration of surgical procedures follows a log-normal distribution and often exceeds the surgeon\textquotesingle s estimates \cite{li09, gomes12}.

Delays in project completion have a negative impact on our society. The project product has an intended economic or social benefit. The materialisation of that benefit will have to wait if the end product is
not delivered on time. There is an additional cost for the project owner as well. An increase in project duration is correlated with an increase in the project cost \cite{emmons07, bromilow69, majerowicz16}.

Understanding what are the causes of project delays can help us anticipate and mitigate their negative effects. In the 80s, Kahneman and Tversky postulated an optimism bias during the planning phase as a key
factor \cite{kahneman79, tversky82}. We tend to underestimate activity durations and, as a consequence, observed activity durations are larger. Kahneman and Tversky proposed looking at past activities of a similar
kind as a corrective measure. A methodology known as reference class forecasting. Nowadays, the theory of Kahneman and Tversky is having a renaissance at the Flyvbjerg school of reference class forecasting \cite{flyvbjerg06}. Given data about a quantity of interest and its associated metadata, reference class forecasting  performs an unsupervised or supervised clustering of the data to determine the existence of groups and the group dependent distribution of the quantity. The groups and their associated distributions are used later to make predictions about future values of the quantity (e.g., past delays stats to predict future delay). 

The theory of optimism bias does not tell us what is the actual cause of delay. To apply reference class forecasting, we need to specify the properties binding activities into the same class. Surveys, literature
mining and statistical analysis of observed data have been used to investigate delay determinants \cite{ma00, assaf06, stepaniak10, lindhard14, durdyev19}. More recently, methods from machine learning and natural language processing are being used to automate this procedure \cite{egwim21, hong21, zachares22, fitzsimmons22}.

Despite this body of work, we do not know what are the patterns of activity delays in human projects. It is not even clear if there are any delay patterns at all, or whether every project is unique. Other areas
suggest there are. The statistics of interevent times between recurrent activity executions is universal \cite{barabasi05, oliveira05, dezso06, vazquez06, jiang13, yan17}. Whether it is letters, emails, phone calls, web access or github commits, the time between two consecutive events has a heavy tail distribution \cite{barabasi05, oliveira05, dezso06, vazquez06, jiang13, yan17}. Here we show that project delays follow a
universal pattern as well, that we call the law of activity delays.

\section{Motivation}
\label{sec:motivation}

The longer an activity takes, the higher the chance something goes wrong. If the planned execution rate is 1 work unit per day but the actual execution rate is $r<1$ work units per day, then the activity completion is delayed in average by $(1-r)\times({\rm planned\ duration})$. In other words, delays are proportional to activity durations. Other factors may be relevant as well. Because delays are proportional to duration, those other factors should contribute in a multiplicative fashion. If $y$ denotes the activity delay (the impact), $f_1$ the activity duration and $f_2$, \ldots, $f_n$ the remaining $n-1$ factors, then $y = f_1\cdots f_n$. This multiplicative equation can be transformed into an additive one by taking the logarithm: $\log y = \sum_i x_i$, where $x_i = \log f_i$.

In practice, we are not aware of all possible delay factors. Suppose we have $k$ known factors, including duration, while $u$ factors remain unknown. Grouping together the contribution of known and unknown factors,
\begin{equation}
\log y = \sum_{1\leq i\leq k} x_i + \Delta x,
\label{logy_motivation}
\end{equation}
where $\Delta x = \sum_{k<i\leq k+u} x_i$. If the unknown factors are modeled as random variables, then the distribution of their sum can be approximated by the normal distribution $N(\Delta x, u\mu_0, u^{1/2}\sigma_0)$, where $\mu_0$ and $\sigma_0$ are the typical mean and standard deviation of the unknown factors. Setting $\mu = u\mu_0$ and $\sigma = u^{1/2}\sigma_0$ we arrive at the expression $\log y = \mu + \sum_{1\leq i\leq k} x_i + \sigma z$, where $z$ is a random variable with the standardized normal distribution $N(z, 0, 1)$. Here $\mu$ and $\sigma$ are the mean and standard deviation of the residual log-delay ($\Delta\log y = \log y -\sum_{1\leq i\leq k} x_i$). They parametrize the unknown.

\section{The law of activity delays}
\label{sec:law}

We define an activity execution delay as the time difference between its actual duration and its planned duration. Following the standard of risk models, activity delays are characterized by a delay likelihood $p$, and a delay impact $y$. Take for example a nuclear reactor meltdown. For every nuclear reactor there is a likelihood of meltdown, the probability that it will take place. When happening, the impact of the meltdown will vary depending on the reactor, what caused the meltdown, the emergency response, etc. Same for delays. There is a likelihood that an activity is delayed and, if delayed, delays can be of different magnitudes. What we call the impact.

The motivation given in the previous section suggest the existence of a risk scale. That scale should apply in the same way to the likelihood and the impact, aside for some considerations about the nature of those variables. Likelihood, delay probability, is constrained to the $(0,1)$ interval. That can be accommodated by a logistic dependency of risk on risk factors. Impact, the magnitude of the delay is measured in a log scale. That means we error by a multiplicative factor. Rather than plus minus. These can bee seen as strong assumptions. Yet, as shown below, they are supported by the data. Based on these arguments we postulate the law of activity delays
\begin{equation}
p(\vec{x};g_0,g_1) = \frac{1}{ 1 + \exp\left( -(g_0 + g_1 \sum_{1\leq i\leq k} x_i) \right)},
\label{px}
\end{equation}
\begin{equation}
\log y(\vec{x};\mu,\sigma) = \mu + \sum_{1\leq i\leq k} x_i + \sigma z,
\label{logy}
\end{equation}
where $z$ is a random variable with a standardized normal distribution and the logistic function in equation (1) enforces the requirement that $0\leq p\leq 1$.

Given data for $\vec{x}$ and $\log y$, we can estimate $\mu$ and $\sigma$ as the mean and variance of $\log y - \sum_{1\leq i\leq k}x_i$. Once that is done, we can calculate $z$ as follows
\begin{equation}
z = \frac{\log y -\mu - \sum_{1\leq i\leq k} x_i}{\sigma}.
\label{z}
\end{equation}
The variable $z$ is the standardized residual. What is left after we have subtracted all what we know. The fact that $z$ should follow a standardized normal distribution can be used to validate or falsify the postulated law of activity delays, Eqs. (\ref{px})-(\ref{logy}). To avoid confusion, bear in mind that $z$ can be understood both as a variable or as data. If we would to generate random delays, we could extract $z$ values from a standardized normal distribution. In turn, if we would like to challenge Eq. (\ref{logy}), we can collect empirical data of $\vec{x}$ and $\log y$ for different activities, substitute the values in Eq. (\ref{z}) and compute the $z$ distribution. If our postulate is correct, that distribution should be a standardized normal distribution.

\begin{figure}[h]
\includegraphics[width=3in]{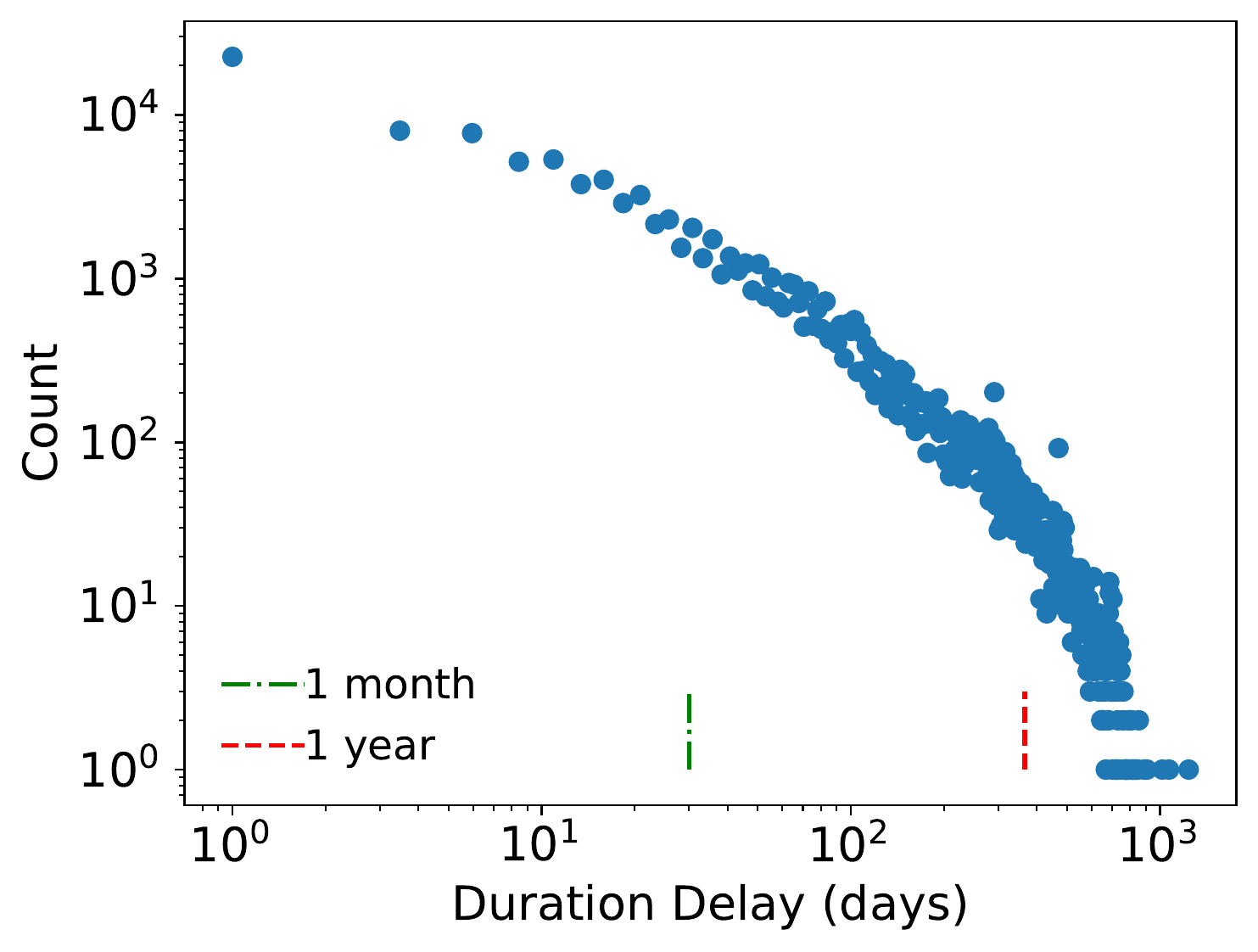}
\caption{The distribution of delays across 108,745 delayed activities from 180 construction projects.}
\label{fig_delay_dis}
\end{figure}

\section{Delay Magnitude}
\label{sec:delay_magnitude}

Before digging into possible delay factors, let's have a look at the distribution of delays without filters (Fig. \ref{fig_delay_dis}). It is evident that delays expand a broad range, from days to years. We could speculate about the shape of this distribution. Yet, there is no point to do so. It represents aggregate data of activities of potentially different types.

\section{Delay factors}
\label{sec:delay-factors}

In this section we uncover delay factors. We gradually increase the number of factors and investigate the convergence of the residual $z$ to the standardized normal distribution. To estimate parameters and validate the model, we pulled together data from 180 construction project schedules (see Appendix \ref{sec:data}). The aggregate dataset has about 300,000 actualized activities, including 100,000 delayed activities. This data was used to estimate the model parameters $(g_0, g_1, \mu, \sigma)$  of the law of activity delays (Eqs. (\ref{px})-(\ref{logy})). The residuals for all delayed activities were then calculated using Eq. (\ref{z}).

\begin{figure}
\includegraphics[width=5.4in]{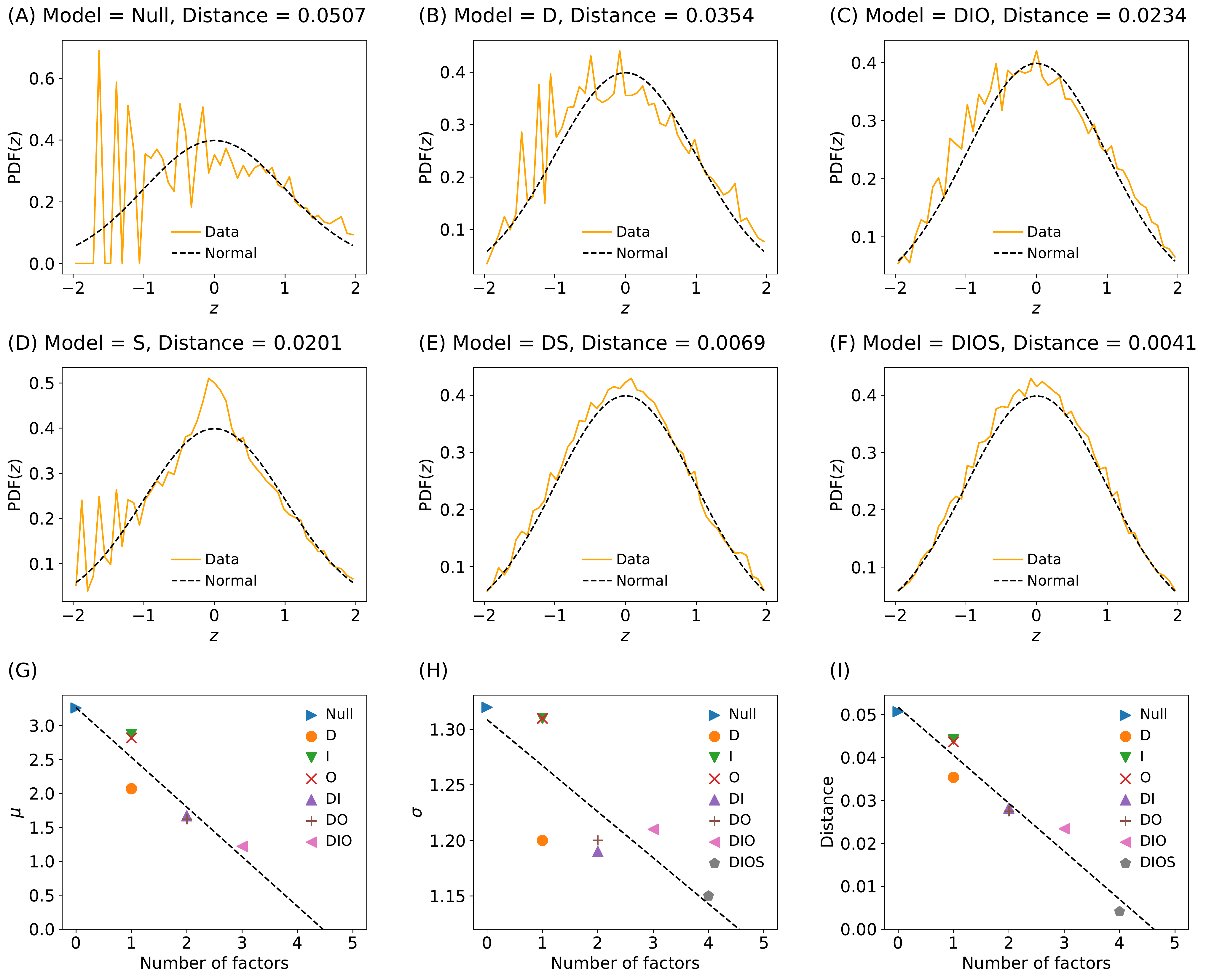}
\caption{The law of activity delays. (A-F) Residuals probability density function for different models (line), based on about 100,000 reported delays across 180 schedules. The dashed line is the postulated standardized normal distribution. (G,H) Residual mean and standard deviation of the log delays versus the numbers of delay factors, a point per model. The dashed line is the best linear regression. (I) Kolmogorov-Smirnov distance between the empirical and the normal distribution versus the number of factors. The Kolmogorov-Smirnov distance is a standard metric to measure distance between distributions. In this case the normal distribution versus the distribution of empirical $z$ values.}
\label{fig_delay_factors}
\end{figure}

\subsection{Null model}

Our starting model, the null model, has no factors: Eqs. (\ref{px})-(\ref{logy}) with $k=0$. As expected, it is not a good model. The distribution of residuals is far from a normal distribution (Fig. \ref{fig_delay_factors}A).

\subsection{Activity duration}

The second model introduces planned activity duration ($D$) as a single delay factor: Eqs. (\ref{px})-(\ref{logy}) with $k=1$ and $x_1=\log D$. The linear regression $\log y = a + b \log D$ results in a slope of $b=1.07$, corroborating our expectation that delays are proportional to activity durations. The model with $x_1=\log D$ exhibits a better agreement between the residuals distribution and the standardized normal distribution (Fig. 1B), relative to what is observed for the null model (Fig 1A), but it is not satisfactory.

\subsection{Activity complexity}

Next, we consider two activity features associated with activity dependencies. The start of an activity is constrained to the completion of $I$ (input) activities it depends on. In turn, a number of $O$ (output) activities are dependent upon the activity completion. The role of activity dependencies in delay propagation along the activity network is well established \cite{santolini21, vazquez23}. A finishing delay of an activity will delay the start or finish of downstream activities depending on the relation type. However, it is not obvious how activity dependencies could increase activity durations. We argue that $I$ and $O$ are indicative of activity complexity and that activity complexity is a risk factor for activity completion delays. The linear regression of $\log y$ vs $\log I$ or $\log O$ results in slopes of 0.84 and 0.81, respectively, corroborating their association with delay. Given they are close to 1 we will assume a slope 1. 

These additions lead to the 3 factor model: Eqs. (\ref{px})-(\ref{logy}) with $k=3$, $x_1=\log D$, $x_2=\log I$  and $x_3=\log O$ (DIO). The DIO model results in a visible improvement in the agreement between the distribution of residuals and the expected standardized normal distribution (Fig. 1C), relative to what is observed with duration alone (Fig \ref{fig_delay_factors}B). Going from D to DIO there is a decrease in the residual log-delays mean $\mu$ (Fig. \ref{fig_delay_factors}G), no significant change in the standard deviation $\sigma$ (Fig. \ref{fig_delay_factors}H) and a decrease in the Kolmogorov-Smirnov distance to the normal distribution (Fig. \ref{fig_delay_factors}I). The factors I and O alone perform as bad as the null model (Fig. \ref{fig_delay_factors}G-I, Table \ref{tab_delay_factors}). From that we conclude, the factors (I,O) are corrections to D.

\begin{table}[t]
\begin{tabular}{l|r|r|r|r|r|r}
\hline
model &   $g_0$ & $g_1$ &       $\mu$ &  $\sigma$ &  KS-$d$ &  KS-$p$ \\
\hline
  Null & -0.923 & 0.000 &  3.26 &   1.32 &        0.051 & 8.111e-204 \\
    D & -1.570 & 0.596 &  2.07 &   1.20 &        0.035 &  1.585e-99 \\
    I & -1.437 & 1.422 &  2.87 &   1.31 &        0.044 & 4.085e-155 \\
    O & -1.532 & 1.522 &  2.82 &   1.31 &        0.044 & 5.934e-152 \\
    S & -1.963 & 0.323 & -0.04 &   1.13 &        0.020 &  2.523e-32 \\
   DI & -1.907 & 0.680 &  1.67 &   1.19 &        0.028 &  2.674e-63 \\
   DO & -1.943 & 0.686 &  1.62 &   1.20 &        0.028 &  3.174e-60 \\
   DS & -2.395 & 0.342 & -1.24 &   1.11 &        0.007 &  3.138e-04 \\
  DIO & -2.191 & 0.686 &  1.22 &   1.21 &        0.023 &  6.590e-44 \\
 DIOS & -2.995 & 0.409 & -2.08 &   1.15 &        0.004 &  9.546e-02 \\
\end{tabular}
\caption{Models fit and performance. The models are constructed with a combination of delay
factors among no factors (Null), activity duration (D), input dependencies (I) and output dependencies (O). We report the Kolmogorov-Smirnov statistics, the distance KS-$d$ between the empirical and expected cumulative distribution function, and the associated statistical significance KS-$p$.}
\label{tab_delay_factors}
\end{table}

\subsubsection{Unknown factors}

The visual inspection of Fig. \ref{fig_delay_factors}C indicates that the empirical distribution of residuals follows the shape of the standardized normal distribution. However, when we perform the Kolmogorov-Smirnov test of normality, we obtain a p-value close to zero, KS-$p=7\times10^{-44}$. Based on that p-value we would reject the null hypothesis that the distribution of residuals is a standardized normal distribution. Therefore, either the postulated law of activity delays is incorrect or we are missing delay factors. We believe it is the latter. As we increase the number of delay factors, the distance KS-$d$ between the empirical and expected distributions decreases and the associated statistical significance KS-$p$ increases (Table \ref{tab_delay_factors}).

In the Motivation section we estimated that $\mu=\mu_0u=\mu_0(n-k)$, where $n$, $k$ and $u$ are the number of total, known and unknown factors determining activity delays. While we do not know $n$, we can plot the estimated $\mu$ for each model as a function of the model number of factors. Using that plot we extrapolate to $\mu = 0$ to obtain an estimate for $n$. Using this approach we estimate that activity delays are determined by 4 to 5 factors. Given we have uncovered 3 factors (D, I, O), that means we have 1 or 2 unknown factors.

\subsubsection{Activity name signatures}
\label{sec:activity-signatures}

Activity names (signatures) contain information about their categories and categories about the risk of delay. For example, outdoor work is susceptible to weather conditions \cite{ballesteros18}. From that observation we infer that the term "outdoor work" in activity signatures is indicative of similar delay statistics. We may have other examples in mind, but in general there is no obvious relationship between activity names and delay risk. Natural Language Processing (NLP) has been proven a powerful method to uncover hidden relationships between text signatures and continuous variables \cite{hong21, zachares22}. Using NLP methods we constructed a map between the activity signature $S$ and a delay factor $x_S$, where $x_S$ is the expected log delay given the signature $S$ (see Appendix \ref{sec:data}). The map is constructed in two steps. First, a mapping of the activity names to a vector of word frequencies (vectorization). Second, a non-linear regression between those vectors and log-delay (Appendix \ref{sec:data}). Off note, the top 5 words indicative of high delay risk are detailed, piping, cutting, excavation, construction. The bottom 5 are revue, redundant, pour, idc, check, where idc stands for inter discipline check. The activity signature factor was then incorporated into the law of delays Eqs. (\ref{px})-(\ref{z})).

First we tested the delay factor $x_S$ alone. The residual distribution distance to the normal is slightly smaller than using activity duration (Fig. \ref{fig_delay_factors}E vs 1B, Table \ref{tab_delay_factors}), and comparable to the DIO model (Fig. \ref{fig_delay_factors}E vs \ref{fig_delay_factors}C). Adding duration, DS model, decreases the distance even further (Fig. \ref{fig_delay_factors}E vs D). The final boost comes when we combine all, the DIOS model (Fig. \ref{fig_delay_factors}F). The Kolmogorov-Smirnov distance between the empirical and the postulated distributions is 5 times smaller for the DIOS than the DIO or S models. Furthermore, the Kolmogorov-Smirnov p-value goes up to 0.095, in the range where we cannot reject the null hypothesis that the residual follows a standardized normal distribution.

In this analysis we mapped activity name to a 100 words vector, taking into account the 100 most frequent words. Although we have about 100,000s data points, we cannot exclude overfitting. To address this we did topic modeling. That means we mapped the activity names to a reduced number of topics, 10 to be more specific. That's a 10 fold reduction in the number of parameters acting on the regression between delays  and the activity signature scores. We did not observe any statistical difference between the unexplained variance $\sigma$ nor the KS distance to the normal distribution, when using the 100 most frequent words or their clustering in 10 topics. From that evidence we conclude there is no evidence for overfitting.

\begin{figure}[t]
\includegraphics[width=5.4in]{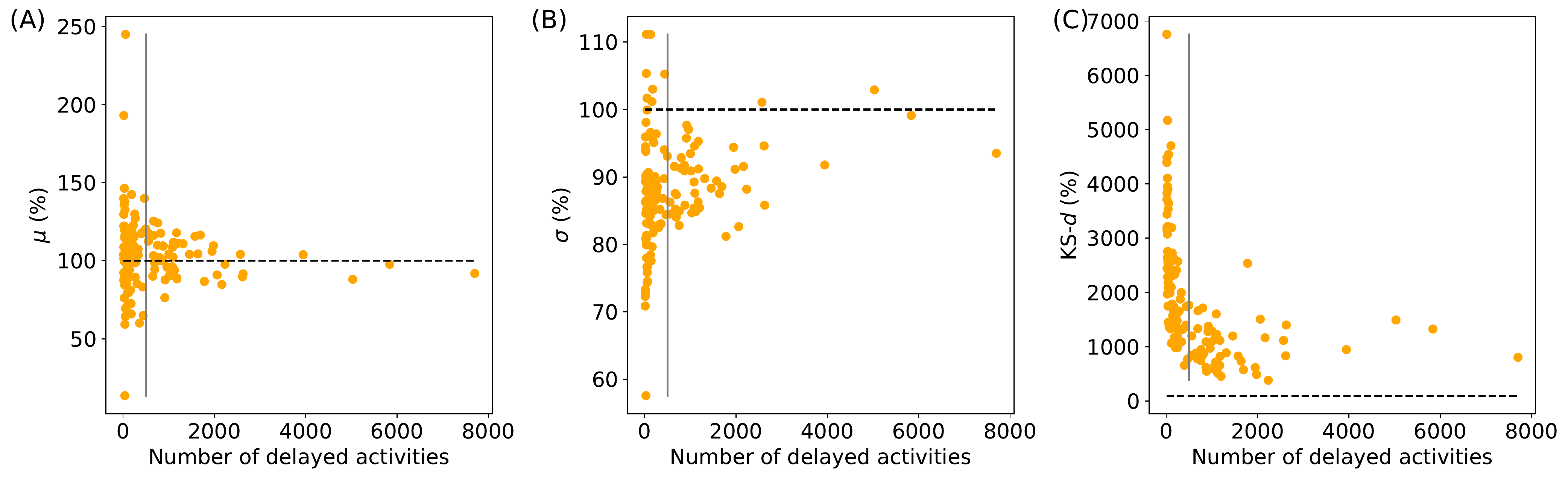}
\caption{Statistics across projects. (A, B) Mean and standard deviation of the non-scaled residual delay ($\log y - \sum_{i=D,I,O,S} x_i$) for each specific project (symbols) as a function of the number of delayed activities, normalized to their values for the aggregate dataset. C) Kolmogorov-Smirnov distance between the distribution of scaled residuals ($z$) and the normal distribution. One point for each of the 180 projects in our database. The DIOS model was trained on the whole dataset excluding a given project and the residuals computed on the given project data. The horizontal dashed line represents the DIOS model values when fitted to the whole dataset.The vertical line marks 500 delayed activities.}
\label{fig_unique}
\end{figure}

\section{Is your project unique?}
\label{sec:unique}

All projects that we have analyzed satisfy the law of activity delays in Eqs. (\ref{px})-(\ref{logy}). To test that observation in a systematic manner, we tested the DIOS model using single project data. That includes training the mapping between activity signatures and delays (the $x_S$ factor). Figure \ref{fig_unique}A and B report the mean and standard deviation of the non-scaled residuals, relative to the values training on the whole dataset. Figure \ref{fig_unique}C reports the Kolmogorov-Smirnov distance between the empirical distribution of scaled residuals and the normal distribution. One point for each of the 180 projects in our database, ordered by the number of actualized (finished) activities they contain.

When there is enough data ($>500$ delays) there is a convergence towards the values for the whole dataset (Fig. \ref{fig_unique}A-C). The symbols for individual projects spread around the dashed line representing the whole dataset. From this analysis we conclude a project is unique to the extent it has a unique distribution of delay factors (D, I, O, S) across activities.

\begin{figure}[t]
\includegraphics[width=3in]{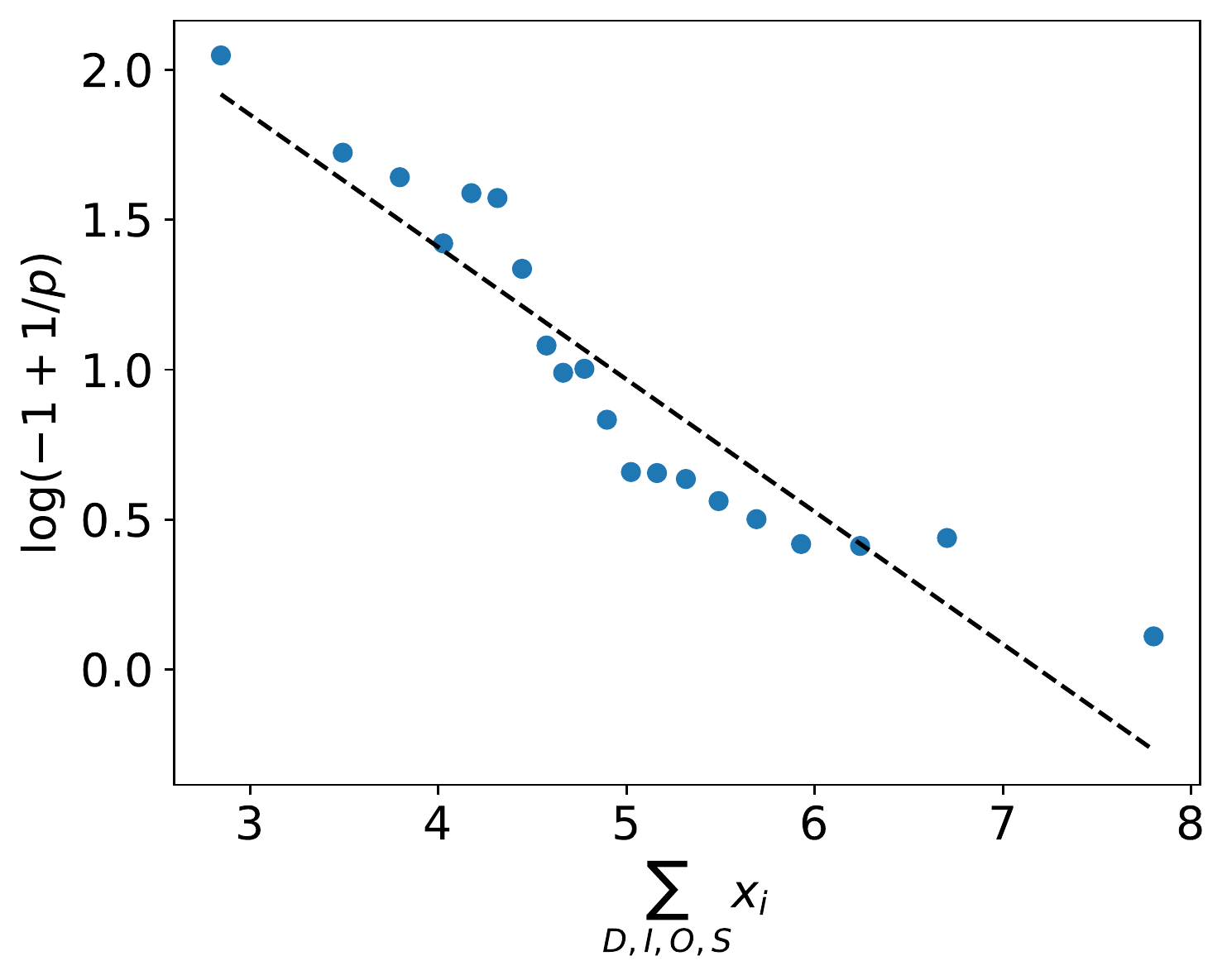}
\caption{Linear scaling between $-1+1/p$ and $\sum_{D,I,O,S}x_i$, as deduced from the postulate in Eq. (\ref{px}). Each point represents an average over }
\label{fig_p_x}
\end{figure}

\section{Scaling of the delay likelihood}

So far we have focused our attention on the delay impact. At this point we will test our postulate about delay likelihood, Eq. (\ref{px}). To this end we calculated $\sum_{D,I,O,S}x_i$ for every activity, binned the activities by their $\sum_{D,I,O,S}x_i$ values and, for each group we estimated the fraction of delayed activities $p$. If the postulated logistic dependency in Eq. (\ref{px}) is correct, then the plot $-1+1/p$ as a function of $\sum_{D,I,O,S}x_i$ should be linear. This is indeed the case, as highlighted by the dashed line in Fig. \ref{fig_p_x}.

\section{Conclusions}
\label{sec:conclusions}

In conclusion, data for 180 construction projects validates our law of activity delays Eqs. (\ref{px})-(\ref{logy}). In practical terms, that means the distribution of x-factors across activities determines the statistics of activity delays. This conclusion has a number of key implications:

Activity delays follow a lognormal distribution with a log mean determined by the delay x-factors.

Two projects are similar or distinct if their distributions of x-factors across activities are similar or distinct. Your project is unique if it has a unique distribution of x-factors across activities.

At the activity level, reference class forecasting is valid to the extent it buckets together activities with similar x-factors. At the project level, to the extent it buckets together projects with similar distributions of x-factors across activities.

Our collection of projects contains different geographical regions, construction companies and periods of time. Those idiosyncrasies are relevant to the extent they contributed to the x-factor distribution across activities.

The DIOS model with activity duration, input dependencies, output dependencies and activity signatures is a validated model for delay risk analysis in the construction industry.

The validity of the law of activity delays remains to be tested in other areas. The delay factors of activity duration, input and output are universal. In contrast, the mapping of activity signatures to delay risk is domain-specific.

A final note. There are projects with a mandatory deadline that do not admit delays. In this context, any contingency leading to delays must be resolved. For example, any big sport event will trigger an array of construction projects and, by the constraint of the opening day, they will be all delivered on time. This is achieved at expenses of extract costs above the planned budget. Other types of rules may apply in these cases \cite{alfi07}.
 
\section*{Acknowledgements}

Nodes \& Links Ltd provided support in the form of salary for Alexei Vazquez, Chrysostomos Marasinou, Georgios Kalogridis and Christos Ellinas, but did not have any additional role in the conceptualization of the study, analysis, decision to publish, or preparation of the manuscript.

\appendix

\section{Data preparation and analysis}
\label{sec:data}

Our dataset consists of 180 schedules containing a total of 320,510 finished activities (actualized) and among those 108,745 delayed activities. We define an activity delay event as an instance where the actual duration exceeded the planned duration by more than 2 days. We used the threshold of 2 days rather than 0 days to exclude annotation errors. If delayed, we define the activity delay as the time difference between its actual duration and its planned duration. For duration as a delay factor we used the planned duration. The number of input and output dependencies is extracted from the activity dependencies reported in the schedules. For the numerical calculations, we used $\log(1+D)$, $\log(1+I)$ and $\log(1+O)$ instead of $\log(D)$, $\log(I)$ and $\log(O)$ to avoid a $\log(0)$ value for activities with annotated duration zero, no input or no output dependencies.

The activity signatures were derived using NLP methods. We trained a neural network with a vocabulary composed by the $W=100$ most frequent words in the activity names (signatures) as a predictor of delay. The network was made of 3 layers (input, hidden, output) with \emph{W} nodes each. The mapping of activity names to vectors was done with the CountVectorize class, python package sklearn.feature\_extraction.text. Training and predictions were made with the MLPRegressor class, python package sklearn.neural\_network.

Estimation of $(g_0, g_1)$. We define the delay event variable $e_j = 1$ if activity $j$ is delayed and 0 otherwise. We perform logistic regression of $e_j$ vs $\sum_{1\leq i\leq k}x_{ij}$, where $x_{ij}$ is the value of x-factor $i$ for activity $j$. From this regression, we estimate $g_0$ and $g_1$ in Eq. (\ref{px}). To this end we use the LogisticRegression class of the python module sklearn.linear\_model.

Estimation of $(\mu, \sigma)$. $\mu$ and $\sigma$ are the mean and standard deviation of the non-scaled residual $\Delta\log y_j = \sum_{1\leq i\leq k}x_{ij}$, where the index $j$ runs over all delayed activities.

KS test. The one sample Kolmogorov-Smirnov test was performed with the
kstest class, python module scipy.stats.

\bibliographystyle{elsarticle-harv}

%\bibliography{risk.bib}

\bibliography{cx.bbl}

\end{document}